\documentclass[aps,prl,preprint,groupedaddress]{revtex4}
\usepackage{graphicx}

\begin{document}


\title{Recursion-transform method on computing the complex resistor network with three arbitrary boundaries }


\author{ Zhi-Zhong Tan  \footnote{E-mail: tanz@ntu.edu.cn ; ~~ tanzzh@163.com}}
\address{ Department of Physics, Nantong University, Nantong 226019, China}

\date{\today}

\begin{abstract}
We perfect the recursion-transform method to be a complete theory, which can derive the general exact resistance between any two nodes in a resistor network with several arbitrary boundaries. As application of the method, we give a profound example to illuminate the usefulness on calculating resistance of a nearly $m\times n$ resistor network with a null resistor and three arbitrary boundaries, which has never been solved before since the Green¡¯s function technique and the Laplacian matrix approach are invalid in this case. Looking for the exact solutions of resistance is important but difficult in the case of the arbitrary boundary since the boundary is a wall or trap which affects the behavior of finite network. For the first time, seven general formulae of resistance between any two nodes in a nearly $m\times n$  resistor network in both finite and infinite cases are given by our theory. In particular, we give eight special cases by reducing one of general formulae to understand its application and meaning.


\pagenumbering{arabic}

\end{abstract}

\pacs{   }

\maketitle


\section{1. Introduction}
The computation of the two-point resistance in networks is a classical problem in circuit theory and graph theory since the German scientist Kirchhoff found the node current law and the circuit voltage law in 1845[1]. Modeling the resistor network has become a basic method to solve a series of complicated problems in the fields of engineering, science and technology. Such as the computation of resistances is relevant to a wide range of problems ranging from classical transport in disordered media [2], first-passage processes [3], lattice Greens functions [4], random walks [5], resistance distance [6,7], to hybrid graph theory[8]. For an up-to-date list of references, see Ref.[9]. Modeling the resistor network can also help to carry on the research of exact finite-size corrections in critical systems [10-15]. However, it is usually very difficult to obtain the exact resistance in a complex resistor networks with arbitrary boundary since the boundary is a wall or trap which affects the behavior of finite network. It requires not only the circuit theory but also the innovation of mathematical theories and methods [16-27]. The construction of and research on the models of resistor networks therefore make sense for theories and applications.

For the explicit computation of two-point resistances in a resistor network, Cserti [16] evaluated the two-point resistance using the lattice Green¡¯s function. His study is confined to regular lattices of infinite size. In 2004 Wu [17] formulated a different approach and derived an expression for the two-point resistance in arbitrary finite and infinite lattices in terms of the eigenvalues and eigenvectors of the Laplacian matrix. The Laplacian analysis has also been extended to impedance networks after a slight modification of the formulation of [18]. The Laplacian method has solved a variety of types of resistance network with the normal boundary, besides the previous cases [17], there are also the recent Tan-Zhou-Yang conjecture [19], and a globe [20]. But the applications of the Laplacian approach require a complete knowledge of the eigenvalues and eigenvectors of the Laplacian straightforward to obtain for regular lattices. However, the actual computation depends crucially on the geometry of the network and, for nonregular lattices such as a network with an arbitrary boundary, it is difficult to solve the eigenvalue problem with arbitrary elements.

In recent years, a new method is created by us [21,22], which is easy to solve the problem of nonregular lattices of cobweb and globe[22-25]. This methods splits the derivation into three parts. The first creates a recursion relation (expressed by matrix) between the current distributions on three successive vertical lines. The second part diagonalizes the matrix relation to produce a recurrence relation involving only variables on the same transverse line. The third part derives a recursion relation between the current distributions on the boundary. Basically, the method reduces the problem from two dimensions to one dimension. Here we refer to this method as the recursion-transform approach ( call it R-T method ).

Using the R-T method to compute the equivalent resistance relies on just one matrix along one direction, and the resistance is expressed by single summation [21-27]. Especially, the R-T method is always valid to calculate the resistance when two opposing boundaries change. For example, the R-T method is used solving a cobweb network in 2013[21, 22], next a globe network [23], and a fen and a cobweb network [24]. very recently a cobweb network with $2r$ boundary is solved [25], and a hammock network with free boundary is solved[26], and a resistor network with an arbitrary boundary is solved by the R-T method [27] .

\begin{figure*}
\begin{center}
\includegraphics[width=9cm,bb=0 0 283 177]{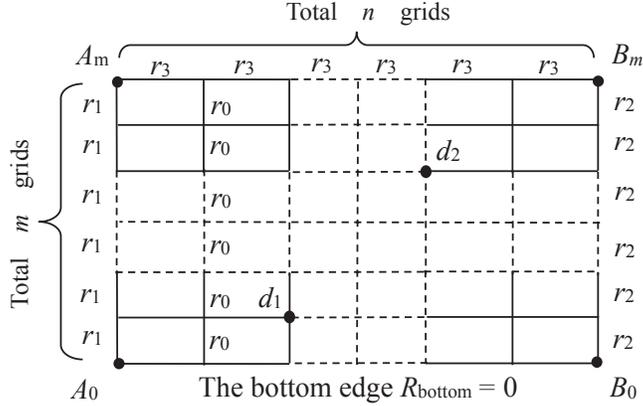}
\caption{ An $m\times n$ resistor network with an null resistor and three arbitrary boundaries.  Bonds in the horizontal and vertical directions are resistors $r$ and $r_{0}$ except for four boundaries.}
\end{center}
\end{figure*}

However the original R-T method is still in its infancy, which has to be perfected. In real life the boundary conditions of the resistor network are often complicated. This paper focus on perfecting the R-T method to be a complete theory, which can derive the general resistance  between any two nodes in the nearly $m\times n$ resistor network with arbitrary boundaries. As applications of the R-T method, we gives three examples to illuminate the usefulness on calculating resistance of a resistor network, and seven general formulae of the resistance between any two nodes in an $m\times n$ resistor network in three cases are given. It shows the R-T method is very interesting and useful to solve the problem which have never been solved before.

\section{2.A general theory and approach}
Fig.1 is called  a nearly rectangular $m\times n$ resistor network with an null resistor and three arbitrary boundaries, where $r$ and $r_{0}$  are, respectively, resistors in the horizontal and vertical directions except for four boundaries, and where $m$ and $n$ are, respectively, the number of meshes along vertical and horizontal directions. Obviously, this model is multipurpose in real life, which has extensive application value. We refer the problem to state the general recursion-transform method.
\subsection{2.1 Overall train of thought and design }
We consider $A_{0}$ is the origin of the rectangular coordinate system, then the left edge act as Y axis and the bottom edge as X axis as shown in Fig.1. Denote nodes of the network by coordinate $(x, y)$.
Assuming the electric current $J$ is constant and goes from the input $d_{1}(x_{1},y_{1})$ to the output $d_{2}(x_{2},y_{2})$. Denote the currents in all segments of the network as shown in Fig.2. The resulting currents passing through all $m+1$ row horizontal resistance of $r$ respectively are: $I_{x,k}^{(1)}, I_{x,k}^{(2)}, I_{x,k}^{(3)}, \cdots I_{x,k}^{(n)}, (0\leq k\leq m)$; the resulting currents passing through all $n+1$ row vertical resistance of $r_{0}$ respectively are:$I_{k}^{(1)}, I_{k}^{(2)}, I_{k}^{(3)}, \cdots I_{k}^{(m)}, (0\leq k\leq n)$.

To find the resistance $ R_{m\times n}(d_{1}, d_{2})$  we can use Ohm's law to  realize it. All nodes on the bottom boundary can be uniformly called O as the resistor of the bottom boundary are all zero. The voltages between the two nodes $O $ and $d_{1}$ , and the two nodes $O $  and $d_{2}$ are respectively
$$ U_{m\times n} (O, d_{1})=r_{0}\sum_{i=1}^{y_{1}}I_{x_{1}}^{(i)} \quad {\text{and}} \quad U_{m\times n} (O , d_{2})=r_{0}\sum_{i=1}^{y_{2}}I_{x_{2}}^{(i)}.   $$
Using Ohm's law, we have
\begin{eqnarray}
{R_{m \times n}}({d_1},{d_2}) = \frac{{{r_0}}}{J}\left( {\sum\limits_{i = 1}^{{y_2}} {I_{{x_2}}^{(i)}}  - \sum\limits_{i = 1}^{{y_1}} {I_{{x_1}}^{(i)}} } \right)
\end{eqnarray}
How to solve the current parameters  $I_{x_{1}}^{(i)}$  and  $I_{x_{2}}^{(i)}$   is the key to the problem.  We are going to solve the problem by modeling matrix equation in terms of Kirchhoff¡¯s law.

\begin{figure*}
\begin{center}
\includegraphics[width=7cm,bb=0 0 219 182]{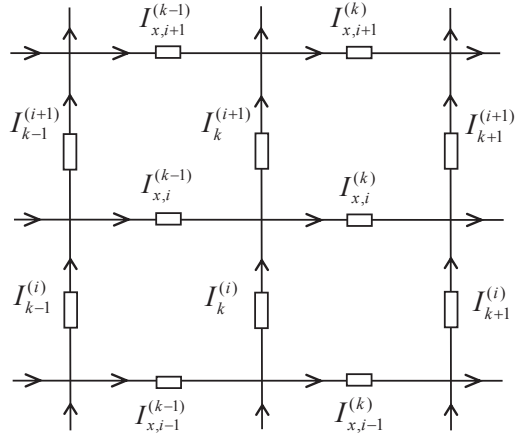}
\caption{ The sub-network model containing the directions of currents  }
\end{center}
\end{figure*}

\subsection{2.2 Modeling the matrix equation}
A segment of the rectangular network is shown in Fig.2. Using Kirchhoff¡¯s law ( KCL and KVL ) to study the resistor network, the nodes current equations and the meshes voltage equations can be achieved from Fig.2. We focus on the four rectangular meshes and nine nodes, this gives the relation,
\begin{eqnarray}
I_{k + 1}^{(1)} &=& (2 + h)I_k^{(1)} - hI_k^{(2)} - I_{k - 1}^{(1)}, \quad i=1 \nonumber\\
I_{k + 1}^{(i)} &=& (2 + 2h)I_k^{(i)} - hI_k^{(i - 1)} - hI_k^{(i + 1)} - I_{k - 1}^{(i)}, \quad  1<i<m \nonumber\\
I_{k + 1}^{(m)} &=& (2 + h+h_{3})I_k^{(m)} - hI_k^{(m - 1)} - I_{k - 1}^{(m)} ,\quad  i=m .
\end{eqnarray}
where  $ h_{k}=r_{k}/r_{0} $  and   $ h =r /r_{0} $.
Equation (2) can be written in a matrix form after considering the input and output current
\begin{eqnarray}
{{\bf{I}}_{k + 1}} = {{\bf{A}}_{m \times m}}{{\bf{I}}_k} - {{\bf{I}}_{k - 1}} - J{{\bf{H}}_x}{\delta _{k,x}},
\end{eqnarray}
where {\bf I}$_{k} $ and {\bf H}$_{x} $ are respectively $m\times 1 $  column matrix,  and reads
\begin{eqnarray}
{\bf I}_{k}=[I_{k}^{(1)}, I_{k}^{(2)}, I_{k}^{(3)} ,\cdots,I_{k}^{(m)}]^{T}, \qquad \qquad \qquad\\
{({H_1})_i} = h( - {\delta _{i,{y_1}}} + {\delta _{i,{y_1} + 1}})\quad {\text{and}} \quad
{({H_2})_i} = h({\delta _{i,{y_2}}} - {\delta _{i,{y_2} + 1}}).
\end{eqnarray}
where ${[ \quad ]^T}$ denotes matrix transposes, and ${({H_k})_i}$ is the elements of ${{\bf{H}}_x}$  with the injection of current $J$ at $d_{1}(x_{1}, y_{1})$  and the exit of current $J$ at $d_{2}(x_{2}, y_{2})$. And
\begin{eqnarray}
{\bf A}_{ m}=
\left( {\begin{array}{ccccc}
   {{2+h}} & {{-h}} &{{0}}& {\cdots}&{{0}} \\
   {{-h}} & {{2+2h}} & {{-h}}  &{\cdots} & {\cdots} \\
   {\vdots}&{\ddots} &{\ddots}&{\ddots}&{\vdots}\\
   {{0}} & {\cdots} &{{-h}} & {{2+2h}} & {{-h}} \\
   {{0}} & {\cdots} &{{0}} & {{-h}} & {{2+h+h_{3}}} \\
\end{array}} \right),
\end{eqnarray}
Next let us consider the boundary conditions of the left and right edges in the network. Applying Kirchoff's laws ({\it KCL} and {\it KVL}) to a single mesh adjacent to each of the left and right boundaries we obtain two matrix equations to model the boundary currents,
\begin{eqnarray}
{{\bf{I}}_1} = [{{\bf{A}}_m} - (2 - {h_1}){\bf{E}}]{{\bf{I}}_0} ~~{\text{and}}~~
{h_1}{{\bf{I}}_0} + {{\bf{I}}_2} = {{\bf{A}}_m}{{\bf{I}}_1}.\qquad \\
{{\bf{I}}_{n - 1}} = [{{\bf{A}}_m} - (2 - {h_2}){\bf{E}}]{{\bf{I}}_n} ~~{\text{and}}~~
{h_2}{{\bf{I}}_n} + {{\bf{I}}_{n - 2}} = {{\bf{A}}_m}{{\bf{I}}_{n - 1}}.
\end{eqnarray}
where $\bf E$ is the $m\times m$ identity matrix, matrix ${\bf A}_{m}$ is given by (6) .

Above Eqs.(3)$\sim $(8) are all equations we need to calculate the equivalent resistance of the $m\times n$ resistor network with three arbitrary boundaries.

\subsection{2.3  Approach of the matrix transformation}
How to solve the matrix equation (3) is the key to the problem. In this section we rebuild a new difference equation and solve Eq.(3) indirectly by means of matrix transformations.

Firstly we consider the solution of (3) in the absence of an injected current, namely, $J = 0$. The eigenvalues  $t_{i}$ $( i=1,2,\cdots, m )$ of ${\bf A}_{m}$  are the $m $ solutions of the equation
\begin{eqnarray}
\det \left| {{{\bf{A}}_m} - {t_i}{\bf{E}}} \right| = 0 .
\end{eqnarray}
Since ${\bf A}_{m}$   is a special tridiagonal matrix which is Hermitian£¬it can be diagonalized by a similarity transformation to yield
\begin{eqnarray}
{{\bf{P}}_m}{{\bf{A}}_m}{({{\bf{P}}_m})^{ - 1}} = {{\bf{T}}_m} ,
\end{eqnarray}
where ${{\bf{T}}_m} = {\rm{diag}}\{ {t_1},{t_2}, \cdots ,{t_m}\} $  is a diagonal matrix with eigenvalues $t_{i}$  of  ${\bf A}_{m}$   in the diagonal and column vectors of ${({{\bf{P}}_m})^{ - 1}}$  are eigenvectors of  ${\bf A}_{m}$ .

To find the exact eigenvalues  $t_{i}$  and the exact eigenvectors of the transformation matrix  ${({{\bf{P}}_m})^{ - 1}}$  is the key to the problem. Normally, it is difficult to find them. However, we can obtain the exact eigenvalues and the exact eigenvectors in the case of  $h_{3}=\{0,h,2h\} $. These will be examples in the following section.

We apply  ${{\bf{P}}_m}$ on the left-hand side of (3) and write
\begin{eqnarray}
{{\bf{P}}_m}{{\bf{I}}_{k + 1}} = {{\bf{P}}_m}{{\bf{A}}_m}{{\bf{I}}_k} - {{\bf{P}}_m}{{\bf{I}}_{k - 1}} - J{{\bf{P}}_m}{{\bf{H}}_x}{\delta _{k,x}}.
\end{eqnarray}
By (10) and (11) we define
\begin{eqnarray}
{{\bf{P}}_m}{\bf{I}}_k^{} = {{\bf{X}}_k} \quad {\text{or}} \quad {{\bf{I}}_k} = {\bf{P}}_m^{ - 1}{{\bf{X}}_k}
\end{eqnarray}
where  ${{\bf{X}}_m}$ is a $m\times 1$  column matrix, and reads
\begin{eqnarray}
{\bf X}_{k}=[X_{k}^{(1)},X_{k}^{(2)},\cdots,X_{k}^{(i)},\cdots,X_{k}^{(m)}]^{T} .
\end{eqnarray}
Thus we obtain a new matrix equation after making use of (11) and (12),
\begin{eqnarray}
{{\bf{X}}_{k + 1}} = {{\bf{T}}_m}{{\bf{X}}_k} - {{\bf{X}}_{k - 1}} - J{{\bf{P}}_m}{{\bf{H}}_x}{\delta _{k,x}}.
\end{eqnarray}
Equation (14) is much simpler than (3) as it is a linear equation easily derived. Once $X_{k}^{(i)}$ is solved that all things are easy. Next we conduct the same matrix transformation  used in (11) to (7) and (8).  Applying  ${\bf{P}}_m$  to (7) and (8) on the left-hand sides, we are led to
\begin{eqnarray}
X_1^{(i)} &=& ({t_i} + {h_1} - 2)X_0^{(i)}  \quad {\text{and}} \quad  {h_1}X_0^{(i)} + X_2^{(i)} = {t_i}X_1^{(i)}.\\
X_{n - 1}^{(i)} &=& ({t_i} + {h_2} - 2)X_n^{(i)}  \quad {\text{and}} \quad  {h_2}X_n^{(i)} + X_{n - 2}^{(i)} = {t_i}X_{n - 1}^{(i)}.
\end{eqnarray}
According to the above equations we therefore obtain $X_{x_{1}}^{(i)}$  and $X_{x_{2}}^{(i)}$  after some algebra and reduction the two solutions needed in our resistance calculation (1).

\subsection{2.4  Derivation of the resistance formula}
According to Eq.(1), the key currents $ I_{{x_1}}^{(i)}$ and $I_{{x_2}}^{(i)} $ must be calculated for deducing the equivalent resistance ${R_{m\times n}}({d_1},{d_2})$. By (12) we have ${{\bf{I}}_k} = {\bf{P}}_m^{ - 1}{{\bf{X}}_k}$, then we  could easy obtain $ I_{{x_1}}^{(i)}$ and $ I_{{x_2}}^{(i)}$ . Finally, substituting  $ I_{{x_1}}^{(i)}$  and $ I_{{x_2}}^{(i)}$   into (1), we therefore obtain the resistance  ${R_{m\times n}}({d_1},{d_2})$.

Obviously, the above method of computation is simple and clear, and is easy to understand, which has good theoretical significance and application value for  the science and technology.
As applications, we give seven new formulae by the following three examples in terms of our R-T method, which have not been solved before.

\section{3. Three application examples}
In order to facilitate the  expression  of the equivalent resistance of resistor network, we define variables $F_{k}^{(i)},\beta_{k,s}^{(i)}, G_{n}^{(i)}, S_{k,i}$, and  $ t_i $  by
\begin{eqnarray}
&& \qquad   F_k^{(i)} = (\lambda_{i}^{k}-\bar{\lambda}_{i}^{k})/(\lambda_{i}-\bar{\lambda}_{i}),
\Delta F_k^{(i)} = F_{k+1}^{(i)} - F_k^{(i)}, \nonumber \\
&&\beta _{k,s}^{(i)} = [\Delta F_{{x_k}}^{(i)} + ({h_1} - 1)\Delta F_{{x_k} - 1}^{(i)}][\Delta F_{n - {x_s}}^{(i)} + ({h_2} - 1)\Delta F_{n - {x_s} - 1}^{(i)}], \nonumber \\
&& \quad  G_n^{(i)} = F_{n + 1}^{(i)} + ({h_2} - 1)({h_1} - 1)F_{n - 1}^{(i)} + ({h_1} + {h_2} - 2)F_n^{(i)}. \\
&& \quad  h_{k} =r_{k}/r_{0} (k=1,2,3), ~ h=r/r_{0};\quad {S_{k,i}} = \sin ({y_k}{\theta _i}), \\
&& \qquad  t_{i}=2(1+h)-2h\cos\theta_{i} \ , \quad (i=1,2.3\cdots , m),
\end{eqnarray}
where $\lambda_{i} \bar{\lambda}_{i}=1$,  and  $\lambda_{i}, \bar{\lambda}_{i}$  are for later uses expressed as
\begin{eqnarray}
{\lambda _i} = 1 + h - h\cos {\theta _i} + \sqrt {{{(1 + h - h\cos {\theta _i})}^2} - 1}
\end{eqnarray}
Note that $\theta_{i}$ is different in different cases,  which  will be determined by the different value of $ h_{3} $.
That is to say $\theta_{i}$   depends on  $ h_{3} $.

\subsection{3.1  Example 1 : The main result in the case of $ h_{3}=0 $ }
When  $ h_{3}=0 $ , from (9) and (10) we have  ${\theta _i} = (i - 1)\pi /m $, $(i=1,2,\cdots , m)$  ( see [23] ) ,   and
\begin{eqnarray}
 {\bf P}_{m}=
\left({\begin{array}{cccc}
   {1/\sqrt{2} } & {1/\sqrt{2} } &{\cdots} & {1/\sqrt{2} } \\
   {\cos(1-\frac{1}{2})\theta_{2}} & {\cos(2-\frac{1}{2})\theta_{2}} & {\cdots} & {\cos(m-\frac{1}{2})\theta_{2}} \\
   {\vdots} & {\vdots} & {\ddots}  &{\vdots} \\
   {\cos(1-\frac{1}{2})\theta_{m}} & {\cos(2-\frac{1}{2})\theta_{m}} & {\cdots} & {\cos(m-\frac{1}{2})\theta_{m}} \\
\end{array}} \right).
\end{eqnarray}
The inverse matrix is ${({{\bf{P}}_m})^{ - 1}} = \frac{2}{m}{[{{\bf{P}}_m}]^T}$,  where ${[ ~ ]^T}$ denotes matrix transpose.

The equivalent resistance between any two nodes
${d_1}({x_1},{y_1})$  and  ${d_2}({x_2},{y_2})$ in the $m\times n$ resistor network can be written as ($ n\geq 1$ )
\begin{eqnarray}
{R_{m \times n}}({d_1},{d_2}) = \frac{{{h_1}{h_2}{{({y_2} - {y_1})}^2}}}{{m[(n - 1){h_1}{h_2} + {h_1} + {h_2}]}}{r_0} \qquad \qquad  \nonumber\\
+ \frac{{{r_0}}}{m}\sum\limits_{i = 2}^m {\frac{{{\beta _{1,1}}S_{1,i}^2 - 2{\beta _{1,2}}{S_{1,i}}{S_{2,i}} + {\beta _{2,2}}S_{2,i}^2}}{{(1 - \cos {\theta _i})G_n^{(i)}}}} ,
\end{eqnarray}
where ${S_{k,i}} = \sin ({y_k}{\theta _i})$, $ {\theta _i} = {(i - 1)\pi}/m $.

In particularly, when $ n\rightarrow \infty,  x_{1}, x_{2}\rightarrow \infty $   with  $x_{1}-x_{2}$   finite, we have result in semi-infinite case,
\begin{eqnarray}
R_{m\times \infty}(d_{1},d_{2})=\frac{r}{m}\sum_{i=2}^{m}\frac{S_1^2 + S_2^2 - 2\bar \lambda _i^{|{x_2} - {x_1}|}{S_1}{S_2}} {\sqrt {{{(1 + h - h\cos {\theta _t})}^2} - 1} } ,
\end{eqnarray}

\subsection{3.2	Example 2 : The main result in the case of  $h_{3}=h$ }
When  $ h_{3}=h $ , from (9) and (10) we have
${\theta _i} = (2i - 1)\pi /(2m + 1)$, $(i=1,2, \cdots , m)$ ( see [24] ), and
\begin{eqnarray}
 {\bf P}_{m}=
\left({\begin{array}{cccc}
   {\cos(1-\frac{1}{2})\theta_{1}} & {\cos(2-\frac{1}{2})\theta_{1}} &{\cdots} & {\cos(m-\frac{1}{2})\theta_{1}} \\
   {\cos(1-\frac{1}{2})\theta_{2}} & {\cos(2-\frac{1}{2})\theta_{2}} & {\cdots} & {\cos(m-\frac{1}{2})\theta_{2}} \\
   {\vdots} & {\vdots} & {\ddots}  &{\vdots} \\
   {\cos(1-\frac{1}{2})\theta_{m}} & {\cos(2-\frac{1}{2})\theta_{m}} & {\cdots} & {\cos(m-\frac{1}{2})\theta_{m}} \\
\end{array}} \right),
\end{eqnarray}
The inverse matrix is  ${({{\bf{P}}_m})^{ - 1}} = \frac{4}{{2m + 1}}{[{{\bf{P}}_m}]^T}$.

The equivalent resistance between any two nodes ${d_1}({x_1},{y_1})$  and  ${d_2}({x_2},{y_2})$  in the $m\times n$ resistor network can be written as ($ n\geq 1$ )
\begin{eqnarray}
R_{m\times n}(d_{1},d_{2})=\frac{2r_{0}}{2m+1}\sum_{i=1}^{m}\frac{\beta _{1,1}^{(i)}S_{1,i}^2 - 2\beta _{1,2}^{(i)}{S_{1,i}}{S_{2,i}} + \beta _{2,2}^{(i)}S_{2,i}^2} {(1 - \cos {\theta _i})G_n^{(i)}} ,
\end{eqnarray}
where ${S_{k,i}} = \sin ({y_k}{\theta _i})$, $ {\theta _i} = {(2i - 1)\pi}/(2m+1) $.

In particularly, when $ n\rightarrow \infty,  x_{1}, x_{2}\rightarrow \infty $   with  $x_{1}-x_{2}$ finite, we have result in semi-infinite case,
\begin{eqnarray}
R_{m\times \infty}(d_{1},d_{2})=\frac{2r}{2m+1}\sum_{i=1}^{m}\frac{S_1^2 + S_2^2 - 2\bar \lambda _i^{|{x_2} - {x_1}|}{S_1}{S_2}} {\sqrt {{{(1 + h - h\cos {\theta _t})}^2} - 1} } ,
\end{eqnarray}

\subsection{3.3	Example 3 : The main result in the case of  $h_{3}=2h $ }
When  $ h_{3}=2h $ , from (9) and (10) we have ${\theta _i} = (2i - 1)\pi /(2m )$, $(i=1,2, \cdots , m)$ ( see [25] ), and
\begin{eqnarray}
 {\bf P}_{m}=
\left({\begin{array}{cccc}
   {\cos(1-\frac{1}{2})\theta_{1}} & {\cos(2-\frac{1}{2})\theta_{1}} &{\cdots} & {\cos(m-\frac{1}{2})\theta_{1}} \\
   {\cos(1-\frac{1}{2})\theta_{2}} & {\cos(2-\frac{1}{2})\theta_{2}} & {\cdots} & {\cos(m-\frac{1}{2})\theta_{2}} \\
   {\vdots} & {\vdots} & {\ddots}  &{\vdots} \\
   {\cos(1-\frac{1}{2})\theta_{m}} & {\cos(2-\frac{1}{2})\theta_{m}} & {\cdots} & {\cos(m-\frac{1}{2})\theta_{m}} \\
\end{array}} \right).
\end{eqnarray}
The inverse matrix is  ${({{\bf{P}}_m})^{ - 1}} = \frac{2}{{m}}{[{{\bf{P}}_m}]^T}$.
Note that although  (24) and (27) have the same structure, but in which the elements  $\theta_{i}$ are different from each other.

The equivalent resistance between any two nodes ${d_1}({x_1},{y_1})$  and ${d_2}({x_2},{y_2})$   in the $m\times n$ resistor network can be written as ($ n\geq 1$ )
\begin{eqnarray}
R_{m\times n}(d_{1},d_{2})=\frac{r_{0}}{m}\sum_{i=1}^{m}\frac{\beta _{1,1}^{(i)}S_{1,i}^2 - 2\beta _{1,2}^{(i)}{S_{1,i}}{S_{2,i}} + \beta _{2,2}^{(i)}S_{2,i}^2} {(1 - \cos {\theta _i})G_n^{(i)}} ,
\end{eqnarray}
where ${S_{k,i}} = \sin ({y_k}{\theta _i})$, $ {\theta _i} = {(2i - 1)\pi}/(2m) $.

In particularly,  when $ n\rightarrow \infty,  x_{1}, x_{2}\rightarrow \infty $   with  $x_{1}-x_{2}$ finite, we have result in semi-infinite case,
\begin{eqnarray}
R_{m\times \infty}(d_{1},d_{2})=\frac{r}{m}\sum_{i=1}^{m}\frac{S_1^2 + S_2^2 - 2\bar \lambda _i^{|{x_2} - {x_1}|}{S_1}{S_2}} {\sqrt {{{(1 + h - h\cos {\theta _t})}^2} - 1} } .
\end{eqnarray}

\subsection{3.4  Example 4 : A general formula of the infinite resistor network}
when $ m, n\rightarrow \infty $, but  $x_{1}-x_{2}$ and $y_{1}-y_{2}$ are finite, by (23), (26) and (29) we convert sums to integral by means of the limit,
\begin{eqnarray}
R_{\infty\times \infty}(d_{1},d_{2})=\frac{r}{\pi}\int_{0}^{\pi}\frac{1 - \bar \lambda _\theta ^{\left| {{x_2} - {x_1}} \right|}\cos ({y_1} - {y_2})\theta } {\sqrt {{{(1 + h - h\cos {\theta})}^2} - 1} } ,
\end{eqnarray}
where  ${\bar \lambda _\theta } = 1 + h - h\cos \theta  - \sqrt {{{(1 + h - h\cos \theta )}^2} - 1} $ .\\

Please note that (30) is from different cases of  $h_{3}=\{0,h,2h \}$. However, we will fully understand the fact since the resistance of the infinite resistor network has nothing to do with the boundary conditions. Thus formula (30) is a general formula of the infinite resistor network with four arbitrary boundaries.

\section{4.  Applications of resistance formula (25) }
The above formulae obtained in this paper are too complicated to understand, we will give several special cases of (25) to understand its application and meaning. When $h_1$  and  $h_2$, $m$ or $n$ are special number values, we have the following special cases.\\
Case 1. When $d_{1}=(x_{1}, 0)$  is at bottom edge, and  $d_{2}=(x_{2}, y_{2})$  is an arbitrary node, from (25) we have
\begin{eqnarray}
R_{m \times n}^{}({d_1},{d_2}) = \frac{{2{r_0}}}{{2m + 1}}\sum\limits_{i = 1}^m {\bigg( {\frac{{\beta _{2,2}^{(i)}}}{{G_n^{(i)}}}} \bigg)\frac{{{{\sin }^2}({y_2}{\theta _i})}}{{1 - \cos {\theta _i}}}} .
\end{eqnarray}
Case 2. When $d_{1}$  and  $d_{2}$  are both on the left edge at  $d_{1}=(0, y_{1})$  and  $d_{2}=(0, y_{2})$, from (25) we have
\begin{eqnarray}
R_{m \times n}^{}({d_1},{d_2}) = \frac{{2{r_1}}}{{2m + 1}}\sum\limits_{i = 1}^m {\bigg( {\frac{{\Delta F_n^{(i)} + ({h_2} - 1)\Delta F_{n - 1}^{(i)}}}{{G_n^{(i)}}}} \bigg)\frac{{{{({S_{1,i}} - {S_{2,i}})}^2}}}{{1 - \cos {\theta _i}}}} .
\end{eqnarray}
Cases 3. When $n\rightarrow\infty $,  with $x_{1}=x_{2}=0 $,  from (25) we have
\begin{eqnarray}
R_{m \times \infty }(\{ 0,{y_1}\} ,\{ 0,{y_2}\} ) = \frac{{2{r_1}}}{{2m + 1}}\sum\limits_{i = 1}^m {\bigg( {\frac{{{\lambda _i} - 1}}{{{\lambda _i} + {h_1} - 1}}} \bigg)\frac{{{{({S_{1,i}} - {S_{2,i}})}^2}}}{{1 - \cos {\theta _i}}}} .
\end{eqnarray}
Cases 4. When $ m, n\rightarrow\infty $,  with $x_{1}=x_{2}=0 $,  from (33) we have
\begin{eqnarray}
R_{\infty \times \infty }(\{ 0,{y_1}\} ,\{ 0,{y_2}\} ) = \frac{{{r_1}}}{\pi }\int_0^\pi  {\frac{{({\lambda _\theta } - 1)[1 - \cos ({y_1} - {y_2})\theta ]}}{{({\lambda _\theta } + {h_1} - 1)(1 - \cos \theta )}}} d\theta .
\end{eqnarray}
Case 5. When $d_{1}=(0, y_{1})$   is at the left edge, and  $d_{2}=(n, y_{2})$  is at the right edge, from (25) we have
\begin{eqnarray}
{R_{m \times n}}(\{ 0,{y_1}\} ,\{ n,{y_2}\} ) = \frac{{2{r_0}}}{{2m + 1}}\sum\limits_{i = 1}^m { {\frac{{{h_1}{\alpha _{2,n}}S_{1,i}^2 + {h_2}{\alpha _{1,n}}S_{2,i}^2 - 2{h_1}{h_2}{S_{1,i}}{S_{2,i}}}}{{(1 - \cos {\theta _i})G_n^{(i)}}}} } ,
\end{eqnarray}
where $\alpha _{k,n}^{(i)} = \Delta F_n^{(i)} + ({h_k} - 1)\Delta F_{n - 1}^{(i)}$ .\\
Case 6. When  $d_{1}$  and  $d_{2}$   are both on the same horizontal axis at $d_{1}=(x_{1}, y)$   and  $d_{2}=(x_{2}, y)$ , from (25) we have
\begin{eqnarray}
{R_{m \times n}}({d_1},{d_2}) = \frac{{2{r_0}}}{{2m + 1}}\sum\limits_{i = 1}^m {\bigg( {\frac{{{\beta _{1,1}} - 2{\beta _{1,2}} + {\beta _{2,2}}}}{{(1 - \cos {\theta _i})G_n^{(i)}}}} \bigg)} {\sin ^2}(y{\theta _i}) .
\end{eqnarray}
Case 7. When $ h_{1}=h_{2}=1 $, the network degrades into a fan network [24], from (25) we have
\begin{eqnarray}
{R_{m \times n}}({d_1},{d_2}) = \frac{{2{r_0}}}{{2m + 1}}\sum\limits_{i = 1}^m {\frac{{\beta _{1,1}^{(i)}S_{1,i}^2 - 2\beta _{1,2}^{(i)}{S_{1,i}}{S_{2,i}} + \beta _{2,2}^{(i)}S_{2,i}^2}}{{(1 - \cos {\theta _i})F_{n + 1}^{(i)}}}} ,
\end{eqnarray}
where   $\beta _{k,s}^{(i)} $ reduces to  $\beta _{k,s}^{(i)} = \Delta F_{{x_k}}^{(i)}\Delta F_{n - {x_s}}^{(i)}$.
Note that the fan network has been resolved by [24] in 2014, formula (37) is completely equivalent to the result in [24]. Obviously the result in [24] is our special case in formula (25).\\
Case 8. When  $h_{2}=0 $, from (25) we have
\begin{eqnarray}
{R_{m \times n}}({d_1},{d_2}) = \frac{{4r}}{{2m + 1}}\sum\limits_{i = 1}^m {\frac{{\beta _{1,1}^{(i)}S_{1,i}^2 - 2\beta _{1,2}^{(i)}{S_{1,i}}{S_{2,i}} + \beta _{2,2}^{(i)}S_{2,i}^2}}{{\Delta F_n^{(i)} + ({h_1} - 1)\Delta F_{n - 1}^{(i)}}}} ,
\end{eqnarray}
where   $\beta _{k, s}^{(i)} $ is defined as $\beta _{k, s}^{(i)} = [\Delta F_{{x_k}}^{(i)} + ({h_1} - 1)\Delta F_{{x_k} - 1}^{(i)}]F_{n - {x_s}}^{(i)}$.

The above eight cases are discovered for the first time, which shows our formulae (22), (25) and (28) have good application value and can deduce a lot of new results.

\section{5.  Summary and discussion}
This paper perfect the infantile R-T method to be a complete theory, which can derive the general resistance between any two nodes in the nearly $m\times n$ resistor network with arbitrary boundaries. As applications of the R-T method, we give three examples to illuminate the usefulness on calculating resistance of a resistor network, and seven general formulae of the resistance between any two nodes in an $m\times n$ resistor network are given by the R-T method, which is mainly composed of the characteristic roots, and is simpler and can be easier to use in practice. Our three examples can't be solved by other methods such as Green¡¯s function technique and the Laplacian approach. This shows the R-T method is very interesting and useful to solve the problem of complicated resistor network.

In particular, our results (22), (25), (28) and (30) are four general formulae, from them we can deduce many simple results. Such as when we take special values of $h_1 = h_2 = 1$, the general formula (25) immediately reduce to result in [24].  Formula (30) is a new general result of the infinite resistor network with arbitrary boundary, which is much more concise and meaningful than the result in [16] and [17] since it is expressed by a single integral rather than a double integral.

In addition, a very important usefulness is that the R-T method can be extended to impedance networks, since the Ohm's law based on which the method is formulated is applicable to impedances. In addition, the grid elements $r$ and $r_{0}$  can be either a resistor or impedance, we can therefore study the $m\times n$ complex impedance networks, such as
\begin{eqnarray*}
r = {\text i}\omega L + \frac{R}{{1 + {\text i}\omega CR}}, \quad
{r_0} = {\text i}\omega {L_0} + \frac{{{R_0}}}{{1 + {\text i}\omega {C_0}{R_0}}}.
\end{eqnarray*}
Based on the plural analysis, the formula of the equivalent impedances of the $m\times n$  $RLC $ network with multivariate boundary can be obtained.

\section*{Acknowledgment}
This work is supported by Jiangsu Province Education Science 12-5 Plan Project (No. D/2013/01/048). We thank Professor F. Y. Wu and Professor J.W. Essam for giving some useful guiding suggestions.

\end{document}